\documentclass[runningheads]{llncs}
\usepackage{graphicx}
\graphicspath{{figs/}}
\usepackage{bm,amsfonts,makecell,multirow,stfloats,booktabs,subfigure,amsmath}
\usepackage[misc]{ifsym}
\begin{document}
\title{Semantic Gaussian Mixture Variational Autoencoder for Sequential Recommendation\thanks{This work was supported by Chongqing Science and Technology Bureau (CSTB2022TIAD-KPX0180) and the National Natural Science Foundation of China
under Grant No. 62072450.}}
\titlerunning{ Semantic Gaussian Mixture VAE for Sequential Recommendation}
%
%
\author{Beibei Li\inst{1} \and Tao Xiang \Letter \inst{1} \and Beihong Jin \Letter \inst{2,3} \and Yiyuan Zheng\inst{2,3} \and Rui Zhao\inst{2,3}}
\authorrunning{B. Li et al.}
\institute{College of Computer Science, Chongqing University, Chongqing, China \email{\{libeibeics,txiang\}}@cqu.edu.cn \and 
Key Laboratory of System Software (Chinese Academy of Sciences) and State Key
Laboratory of Computer Science, Institute of Software, Chinese Academy of Science, Beijing, China \email{Beihong@iscas.ac.cn} \and University of Chinese Academy of Sciences, Beijing, China }

%
\maketitle              
\begin{abstract}


Variational AutoEncoder (VAE) for Sequential Recommendation (SR), which learns a continuous distribution for each user-item interaction sequence rather than a determinate embedding, is robust against data deficiency and achieves significant performance. However, existing VAE-based SR models assume a unimodal Gaussian distribution as the prior distribution of sequence representations, leading to restricted capability to capture complex user interests and limiting recommendation performance when users have more than one interest.  Due to that it is common for users to have multiple disparate interests, we argue that it is more reasonable to establish a multimodal prior distribution in SR scenarios instead of a unimodal one. Therefore, in this paper, we propose a novel VAE-based SR model named SIGMA. SIGMA assumes that the prior of sequence representation conforms to a Gaussian mixture distribution, where each component of the distribution semantically corresponds to one of a user’s multiple interests. For multi-interest elicitation, SIGMA includes a probabilistic multi-interest extraction module that learns a unimodal Gaussian distribution for each interest according to implicit item hyper-categories. Additionally, to incorporate the multimodal interests into sequence representation learning, SIGMA constructs a multi-interest-aware ELBO, which is compatible with the Gaussian mixture prior.  Extensive experiments on public datasets demonstrate the effectiveness of SIGMA. The code is available at https://github.com/libeibei95/SIGMA.

 acceptance.

\end{abstract}

\section{Introduction}
Sequential Recommendation (SR) \cite{GRU4Rec,GRURec+} aims to predict the next item to be interacted with according to a user's history interaction sequence. By learning the evolving trends of user interests over time and the transition between items, SR models can capture user interests and predict target items accurately.  In recent years, deep SR models have incorporated various deep neural networks (DNNs) such as recurrent neural networks, convolutional neural networks, and attention mechanisms. These deep SR models encode interaction sequences into embeddings with DNNs, then calculate prediction scores according to the similarities between sequences and candidate items, and finally select the items with the highest prediction scores as recommendations. Among existing deep SR models, the attention mechanisms, typified by the Transformer \cite{Transforemer}, have shown promising performance by leveraging their advantages in modeling long-distance dependencies \cite{STAMP,SASRec}.  Recent Transformer-based SR models integrate with other techniques, such as contrastive learning \cite{CL4SRec,PDMRec},  to further enhance recommendation performance.

However, most of existing SR models learn determinate representations for sequences, which lack smoothness, i.e., small perturbations of embeddings can lead to totally different recommendations.  Therefore, they are easy to be deteriorated by the common data sparsity and data noise problems, which disturb sequence embedding learning and have not been well-solved. To address this issue, Variantial AutoEncoder (VAE) \cite{VAE,VAE2}, a representative probabilistic latent variable model, is introduced into SR \cite{SVAE,VSAN} and represents sequences with distributions rather than embeddings, so that the representation continuity is guaranteed. VAE-based SR models combine the strong representation power of latent spaces provided by VAE, with the sequence modeling capabilities of DNNs, achieving significant performance improvement. 

Unfortunately, conventional VAE suffers from posterior collapse caused by over-regularization \cite{wu2018multimodal}, i.e., the estimates posterior distributions of different input data are undistinguishable in the latent space due to that they are all optimized towards the standard Gaussian distribution. VAE-based SR models \cite{SVAE}  inherit this problem. To resolve the problem, adversarial training and contrastive learning\cite{ACVAE,ContrastVAE} are introduced into VAE-based SR,  enhancing the quality of latent variables and achieving substantial performance improvement. Nevertheless, these models assume that sequence representations conform to the unimodal prior distributions, which exhibit limitations in capturing complex and diverse user interests. In real-world SR scenarios, it is common for a user's interaction sequence to involve multiple interests. For instance, on an e-commerce platform, a user may simultaneously browse  clothes, skin care products, and kitchenware, which reflects his/her multiple interests. Therefore, it is more reasonable to assume that the sequence representations are generated from a multimodal latent space rather than a unimodal one.

Multi-interest recommendation models \cite{cen_controllable_2020,CMI}, which reveal multiple interests implied in history interactions and learn one embedding for each interest, explore multimodal latent space. Compared with SR models that learn one embedding from each interaction sequence, multi-interest recommendation models disentangle multiple embeddings with different semantics from each sequence, which is a much harder task and is more susceptible to data sparsity and noise. Moreover, similar to most SR models, existing multi-interest recommendation models also learn determinate embeddings rather than distributions for interests, which further amplifies the influence of data deficiency. 

To enhance the representation capability of the latent space in VAE-based SR, we propose a novel VAE model named \textbf{SIGMA} (\textbf{S}emant\textbf{I}c \textbf{G}aussian \textbf{M}ixture variational \textbf{A}utoencoder) in this paper. Unlike existing models, SIGMA assumes the prior distribution of sequence representation is multimodal and  follows a semantic Gaussian mixture distribution, where each component corresponds to one of a user's multiple user interests. Specifically, SIGMA establishes an Evidence Lower BOund (ELBO) that is compatible with the mixture Gaussian prior for SR. Then, to estimate the distribution of each user interest, SIGMA incorporates a VAE-based multi-interest extraction module, which disentangles the user's multiple interests from the interaction sequence and learns a unimodal Gaussian distribution for each individual interest.

Overall, our contributions are summarized as follows: 

\begin{enumerate}
     \item We propose a novel VAE-based SR model SIGMA that aligns the estimated posterior of sequence representation with a semantic Gaussian mixture distribution, taking into consideration that a user-item interaction sequence reflects the user’s multiple interests.
     \item We disentangle multiple heterogeneous user interests from sequences, with the guidance of orthogonal categories of items and represent each interest with a unimodal Gaussian distribution.
     \item We conduct extensive experiments on three public datasets and demonstrate that the SIGMA outperforms both representative SR models and multi-interest recommendation models.
 \end{enumerate}

\section{Preliminaries}
\vspace{-0.2cm}\subsection{Problem Definition}

We denote the user set as $\mathcal{U}$, the item set as $\mathcal{V}$, and the embedding of an item $v \in \mathcal{V}$ as $\bm{v} \in \mathbb{R}^d$, where $d$ is the dimension of embeddings. For a user $u \in \mathcal{U}$, we sort his/her interactions in ascending order by timestamp, obtaining an interaction sequence $\bm{s}=\left[v_{1}, v_{2}, \ldots, v_{T}\right]$. Our goal is to build a model that predicts the next item that the user $u_i$ is most likely to interact with at the $T+1$ step among the item set $\mathcal{V}$ given the sequence $\bm{s}$ as input:
\begin{equation}
\nonumber
\operatorname{argmax}_{v_j \in \mathcal{V}} P\left(v_{T+1}=v \mid \bm{s}\right).
\end{equation}

\vspace{-0.2cm} \subsection{ELBO for SR}
Given a user interaction sequence $\bm{s}=\left[v_{1}, v_{2}, \ldots, v_{T}\right]$, where $T$ is the length of the interaction sequence, we assume its generative process is as follows:
\begin{equation}
\nonumber
\prod\nolimits_{t=0}^{T-1}p(v_{t+1}\mid \bm{z}_t) = \prod\nolimits_{t=0}^{T-1}p_{{\theta^\prime}}(v_{t+1} \mid \bm{z}_t)p(\bm{z}_t),
\end{equation}
 where $\bm{z}_t\in\mathbb{R}^d$ is the latent variable that generates the next item $v_{t+1}$, $p(\bm{z}_t)$ represents the prior probability of $\bm{z}_t$, and $p_{{\theta^\prime}}(v_{t+1}\mid \bm{z}_t)$ represents the likelihood parameterized by ${\theta^\prime}$. The log-likelihood of the user sequence is expressed as follows:
\begin{align}
\nonumber
\log \prod\nolimits_{t=1}^{T}p(v_{t}) & = \sum\nolimits_{t=1}^{T}\log p(v_{t}) = \sum\nolimits_{t=0}^{T-1}\log\int_{\bm{z}_t}p(\bm{z}_t)p_{\theta^\prime}(v_{t+1}|\bm{z}_t)d\bm{z}_t.
\end{align}

Maximizing the above logarithmic likelihood is intractable since it is necessary to calculate all possible latent variables $\bm{z}$. To address this issue, VAE-based SR models follow the standard VAE to leverage variational inference and optimize the model by maximizing the evidence lower bound (ELBO), which consists of a reconstruction term and a KL-divergence term. Therefore, the ELBO for VAE-based SR can be written in the following form:
\begin{align}
\label{eq:lelbo}
\nonumber
ELBO :=  &  \sum\nolimits_{t=0}^{T-1}\large(\underbrace{\mathbb{E}_{q_{\phi^\prime}(\bm{z}_t \mid \bm{s}_{1:t})}\left[\log p_{\theta^\prime}(v_{t+1}\mid \bm{z}_t)\right]}_{\textit{Reconstruction term } \mathcal{T}_{recon}} -\underbrace{D_{K L}\left[q_{\phi^\prime}(\bm{z}_t \mid \bm{s}_{1:t}) \lVert p(\bm{z}_t)\right]}_{\textit{KL-divergence term }\mathcal{T}_{KL}}\large),
\end{align}
where $\mathcal{T}_{recon}$ and $\mathcal{T}_{KL}$ denote the reconstruction term and KL-divergence term, respectively. $\bm{s}_{1:t}$  represents the prefix subsequence consisting of the first $t$ items of sequence $\bm{s}$, $q_{\phi^\prime}(\bm{z}_t\mid\bm{s}_{1:t})$ is the estimated posterior probability, and $q_{\phi^\prime}(\cdot)$ and $p_{\theta^\prime}(\cdot)$ are the encoder and decoder built on neural networks, respectively. In SR scenarios, the encoder and decoder are usually instantiated with sequence modeling methods such as RNN and Transformer. 

\begin{figure}[!t]
    \centering
    \includegraphics[width=0.7\textwidth]{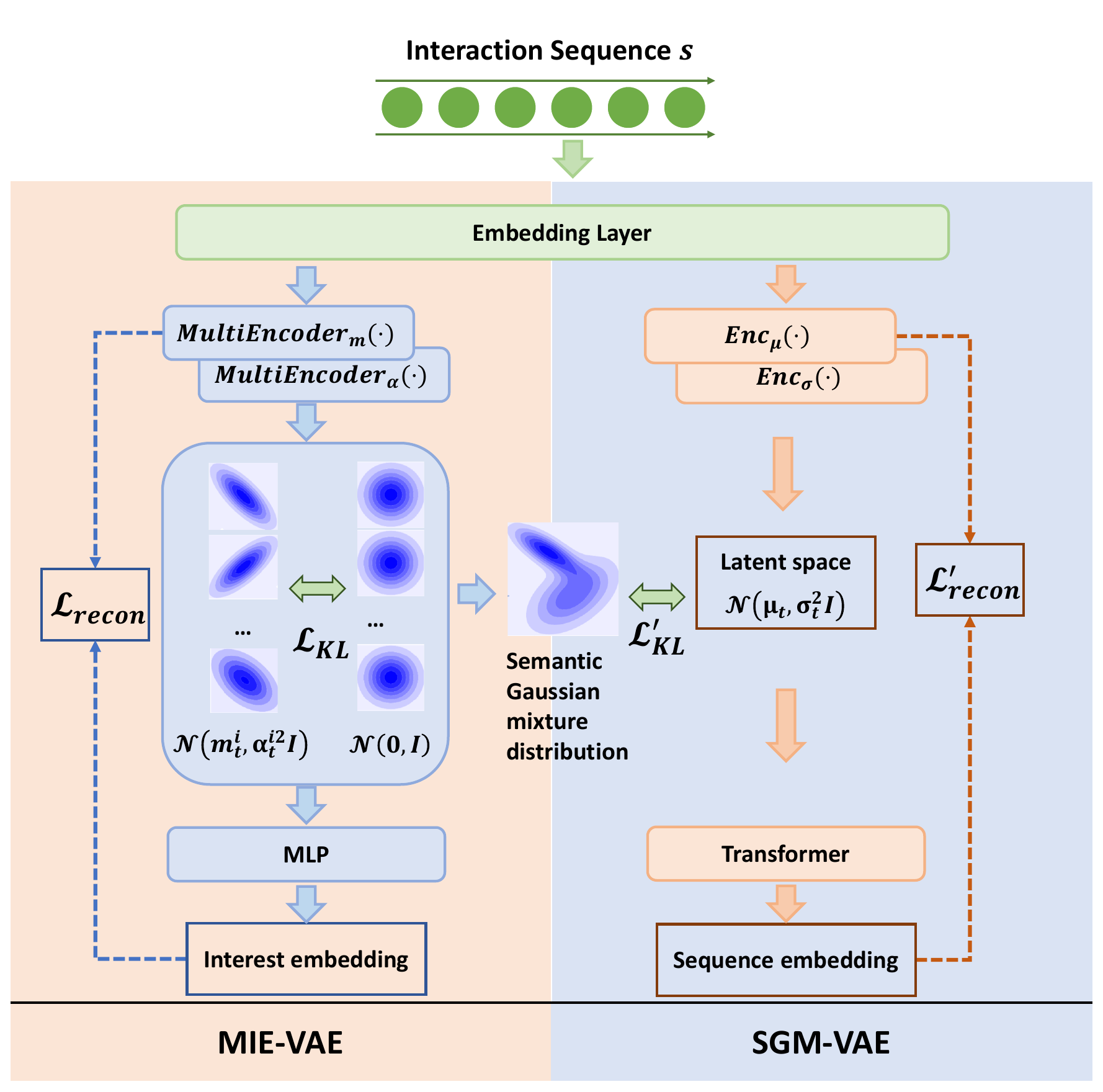} 
    \caption{Architecture of SIGMA. SIGMA comprises of MIE-VAE and  SGM-VAE. MIE-VAE aims to disentangle multiple interests and learn a unimodal Gaussian distribution for each of them, while SGM-VAE learns enhanced sequence representation by aligning with the semantic  Gaussian mixture distribution composed of multiple interests.} 
    \label{fig:model} 
\end{figure}

\section{Methodology}
\vspace{-0.2cm} \subsection{Overview}


As shown in Figure \ref{fig:model}, SIGMA consists of a Multi-Interest Extraction VAE (MIE-VAE) and a Semantic Gaussian Mixture VAE (SGM-VAE). The former focuses on interest representation learning, while the latter is designed for sequence representation learning under the assumption of multi-modal prior distribution. Specifically, MIE-VAE aims to disentangle multiple interests from each sequence, represent each interest with a unimodal Gaussian distribution and quantify the intensity of each interest. The weighted mixture of these interest distributions forms a semantic Gaussian mixture distribution and serves as the prior distribution for sequence representation. SGM-VAE learns the posterior distribution of the sequence representation and aligns it with the Gaussian mixture distribution provided by MIE-VAE using KL divergence.

\vspace{-0.2cm} \subsection{Multi-Interest Extraction VAE (MIE-VAE)} \label{sec:mie}

MIE-VAE aims to extract multiple interest representations $\mathcal{X}_t$ from the user sequence $\bm{s}_{1:t}$, where $\mathcal{X}_t = [\bm{x}_t^1, \bm{x}_t^2, \cdots, \bm{x}_t^k]$ denotes the $k$ user interests extracted from the prefix subsequence $\boldsymbol{s}_{1:t}$,  $\bm{x}_t^i$ follows a unimodal Gaussian distribution, that is, $\bm{x}_t^i\sim\mathcal{N}(\bm{m}_t^i, {\bm{\omega}_t^{i}}^2\bm{I})$, $\bm{m}_t^i$ and $\bm{\omega}_t^{i}$ respectively denote the mean and standard variance of the $i$-th interest, and $\bm{I} \in \mathbb{R}^{k\times k}$ is an identity matrix. The representation of the $i$-th interest, i.e. $\bm{x}_t^i$, is sampled from the distribution $\mathcal{N}\left(\bm{m}_t^i, {\bm{\omega}_t^{i}}^2 \bm{I}\right)$, that is, $\bm{x}_t^i =  \bm{m}_t^i + \bm{\omega}_t^i\odot\bm{\epsilon}$.

We construct an encoder to estimate the posterior probability of interest representations given the sequence $\bm{s}_{1:t}$, as follows:
\begin{align}
\nonumber
q_\phi(\mathcal{X}_t \mid \bm{s}_{1:t}) & =q_\phi\left([\bm{x}_t^1, \bm{x}_t^2, \cdots, \bm{x}_t^k] \mid \bm{s}_{1:t}\right) \\
& =\prod\nolimits_{i=1}^{k} q_\phi\left(\bm{x}_t^i \mid \bm{s}_{1:t}\right) \sim \prod\nolimits_{i=1}^{k} \mathcal{N}\left(\bm{m}_t^i, {\bm{\omega}_t^{i}}^2 \bm{I}\right),
\\
[\bm{m}_t^1, \bm{m}_t^2, \ldots, \bm{m}_t^k] & = \operatorname{MultiEncoder}_{m}(\bm{s}_{1:t}),\\
[\bm{\omega}_t^1, \bm{\omega}_t^2, \ldots, \bm{\omega}_t^k] & = \operatorname{MultiEncoder}_{\omega}(\bm{s}_{1:t}),
\end{align}
where   $\phi$ denotes the parameters in the constructed encoder. In addition, $\operatorname{MultiEncoder}_{m}(\cdot)$ and $\operatorname{MultiEncoder}_{\omega}(\cdot)$ are encoders constructed to compute the mean and standard deviation to represent interests, which are introduced as follows. 

\vspace{-0.2cm}\subsubsection{Multi-Interest Encoder} Items can be divided into several categories, such as health, education, etc. Each category corresponds to one user interest. Therefore, we rely on implicit item categories to learn user interests. 

To mine the item categories, we set up $k$ global category embeddings $\bm{G} = [\bm{g}_1, \bm{g}_2, \ldots, \bm{g}_k]$, where $\bm{g}_*\in{\mathbb{R}}^d$ and 2-norm normalized. To reduce redundancy among different categories, we ensure that the global category embeddings are pairwise orthogonal. Therefore, we construct the following orthogonal constraint loss:
\begin{equation}
\mathcal{L}_{orth}=\sum\nolimits_{i=1}^{k} \sum\nolimits_{j=1, j \neq i}^{k} \bm{g}_{i}^{T} \bm{g}_{j}.
\end{equation}

We calculate the correlation score between $v_i$ and the $k$-th category using $\bm{g}_{k}^{T} \bm{v}_{i}$, where $\bm{v}_{i}$ is normalized to the 2-norm. Further, we utilize correlation scores to compute the classification probability of items into each category via softmax function. For example, the probability of $v_{i}$ belonging to the $j$-th category can be calculated as $a_{i}^j=\frac{\exp \left(\bm{g}_{j}^{T} \bm{v}_{i} / \tau\right)}{\sum\nolimits_{l=1}^{k} \exp \left(\bm{g}_{l}^{T} \bm{v}_{i} / \tau\right)}$, where $\tau$ is a temperature coefficient. The larger $\tau$, the smoother the probability values. 


With the classification probabilities, we learn both \textbf{soft interests} and \textbf{hard interests} via two different strategies, where soft interest allows each item to be related to multiple user interests, while hard interests restrict each item to be exclusively related to only one user interest.

For interaction sequence $\bm{s}_{1:t}$, the $j$-th soft interests $\bm{h}_t^j$ is calculated with the weighted summation of embeddings of his/her interacted items classified into the category $i$, as $\bm{h}_{t}^{j}=\sum\nolimits_{i=0}^{t} a_{i}^j \bm{v}_{i}$. The intensity of the $j$-th interest implied in sequence $\bm{s}_{1:t}$ is calculated as $\alpha_t^j = \frac{\sum\nolimits_{i=1}^{t} a_{i}^j}{\sum\nolimits_{i=1}^{t}\sum\nolimits_{j=1}^k a_{i}^j}$. Interest intensity is employed as the weight of the corresponding interest distribution in the semantic Gaussian mixture prior, which will be illustrated in Section \ref{sec:SGM-VAE}.  

To complement that soft interests cannot capture the evolution of user interests over time, we also learn hard interests for users. We initially assign each interacted item  to a category exclusively and then model the evolution of each user interest over time using sequence models. Naturally, we are supposed to assign each interacted item $v_{i}$ to the category $l$ with the highest classification probability, i.e., $l=\operatorname{argmax}_{1\leq j\leq k}\left(\left\{a_{i j}\right\}\right)$. However, the $\operatorname{argmax}(\cdot)$ operation, although widely used, presents a challenge in training models through gradient back-propagation due to its lack of continuous differentiability. In order to overcome this limitation, we transform the non-differentiable sampling process into a continuous and differentiable one via the Gumbel-Softmax trick \cite{gumbel-softmax}.  According to the hard allocation strategy, the interaction sequence $\bm{s}$ is split into $k$ disjoint subsequences $s_j$, where $1\leq j \leq k$, $\sum\nolimits_{j=1}^{k}\left|\bm{s}_{j}\right|=\left|\bm{s}\right|$. The number of subsequences with non-zero length is the number of hard interests, which is dynamic and personalized. Furthermore, to model the temporal evolution of each user's interest, for each non-empty subsequence, we utilize the sequential model to capture the changes in user interest and generate the corresponding hard interest vector. Here, we leverage GRU \cite{GRU4Rec}. The $i$-th hard interest associated with sequence $\bm{s}_{1:t}$ can be denoted as $\bm{r}_t^i$. 


By combining the soft interests and hard interests mentioned above, we compute the mean vector of the $i$-th interest for user $u$, represented as:
\begin{equation}
    \bm{m}_{t}^i =  (\bm{h}_t^i + \bm{r}_t^i)/2.
\end{equation}

As for the standard deviation $\bm{\omega}_i$ of the $i$-th interest, we employ multi-layer perceptrons with the concatenation of the hyper-parameter embedding $\bm{g}_i$ and the mean vector of user interests $\bm{m}_t^i$ as input, and calculate it as follows:
\begin{equation}
\bm{\omega}_{t}^i =  \operatorname{MLP}([\bm{g}_i \lVert \bm{m}_{t}^i]).
\end{equation}


\vspace{-0.2cm}\subsubsection{ELBO for Multi-interest Recommendation} For the KL-divergence term, we assume the prior $p(\mathcal{X}_t) = \prod\nolimits_{i=1}^kp(\bm{x}_t^i)$ and $p(\bm{x}_t^i)\sim\mathcal{N}(\bm{0}, \bm{I})$ for all interests. Combining with the estimated posterior for $i$-th interest of user $u$ is another Gaussian distribution $\mathcal{N}(\bm{m}_t^i, {\bm{\omega}_t^{i}}^2\bm{I})$, the closed-form solution of the KL-divergence could be easily computed through:
\begin{align}
\nonumber
\mathcal{T}_{K L}^{MIE} & = \sum\nolimits_{t=1}^TD_{K L}[q_\phi(\mathcal{X}_{t} \mid \bm{s}_{1:t}) \lVert p_{\theta}(\mathcal{X}_{t})] \\
\nonumber
& =\sum\nolimits_{t=1}^T\sum\nolimits_{i=1}^k D_{K L}\left[q_\phi\left(\bm{x}_{t}^i \mid \bm{s}_{1:t}\right) \lVert p_{\theta}\left(\bm{x}_i\right)\right] \\
& =\sum\nolimits_{t=1}^T\sum\nolimits_{i=1}^k \sum\nolimits_{j=1}^d\left(\omega^{i2}_{{t}, j}+m_{t, j}^{i2}-1-\log \omega_{t, j}^{i2}\right).
\end{align}

For the reconstruction loss, similar to the previous section, we construct it according to the next-item prediction task, as follows:
\begin{equation}
\mathcal{T}_{recon}^{MIE} = \sum\nolimits_{t=1}^T\mathbb{E}_{q_\phi(\mathcal{X}_{t} \mid \bm{s}_{1:t})} \log [p_{\theta}(v_{(t+1)} \mid \mathcal{X}_{t})].
\end{equation}

We map the $j$-th interest embedding extracted from $\bm{s}_{1:t}$, i.e.,  $\bm{x}_t^j$ sampled from $\mathcal{N}\left(\bm{m}_t^j, {\bm{\omega}_t^{j}}^2 \bm{I}\right)$ to the same space as the item embeddings using two-layer perceptrons, i.e., $\bm{u}_t^j=\operatorname{MLP} ({\bm{x}_t^j})$. For a candidate item $v\in\mathcal{V}$, we calculate its relevance scores with each user's interest. For example, the relevance score with the $i$-th interest is $\bm{u}_t^{jT}\bm{v}$. Following previous works on multi-interest recommendation \cite{MIND}, we consider the maximum relevance score between $v$ and the reconstructed interest representation as the final prediction score. We calculate the interaction probability as follows:
\begin{equation}
p(v_{T+1} = v)= \frac{e^{\max \left(\left\{<\bm{u}_{T}^{j}, \bm{v}> / \epsilon \mid 1\leq j\leq k\right\}\right)}}{\sum\nolimits_{v^\prime \in \mathcal{V}} e^ {\max \left(\left\{<\bm{u}_{T}^{j}, \bm{v}^\prime> / \epsilon \mid 1\leq j\leq k\right\}\right)}}.
\end{equation}

Note that we learn user interest distribution on each $t$, which is equivalent to augmenting training data by truncating prefix sequences. While  benefiting representation learning, these interests can be calculated in parallel via optimized matrix calculations, resulting in no additional time consumption.

\vspace{-0.2cm} \subsection{Semantic Gaussian Mixture VAE (SGM-VAE)}\label{sec:SGM-VAE}

Ideal representations of interaction sequences are supposed to reflect the multiple interests of users. In SGM-VAE, we aim to capture complex and diverse user interests by establishing a multimodal prior distribution for sequence representation. To achieve this, we establish an ELBO compatible with semantic Gaussian mixture distribution, which consists of two components: a KL-divergence term and a reconstruction loss.

\vspace{-0.2cm}\subsubsection{Semantic Gaussian Mixture KL-divergence} Considering that users usually have multiple interests and each interest can be represented by a unimodal Gaussian distribution, naturally, the prior distribution of sequence representation can be assumed to follow Gaussian mixture distributions composed of multiple interests. Given the distribution of interests learned via  MIE-VAE from an interaction sequence $\bm{s}$, the prior distribution of the sequence representation can be represented as the following  Gaussian mixture distribution:
\begin{equation}
p(\bm{z}_t) = \sum\nolimits_{i=1}^k \alpha_t^i\mathcal{N}(\bm{m}_t^i, {\bm{\omega}_t^{i}}^2\bm{I}),
\end{equation}
where $\alpha_t^i$ represents the intensity of the $i$-th interest implied by the sequence $\bm{s}_{1:t}$. Since the representations of different interests are personalized and convey different semantics, we can depend on these interests to obtain a distinct semantic prior for each interaction sequence so as to avoid posterior collapse.

The posterior probability distribution of the latent variable $\bm{z}_t$ for sequence $\bm{s}_{1:t}$ can be represented as:
\begin{equation}
 q_{\phi^\prime}(\bm{z}_t\mid \bm{s}_{1:t})\sim\mathcal{N}(\bm{\mu}_t, \bm{\sigma}_t^2\bm{I}), 
\end{equation}
where $\bm{\mu}_t = Enc_{\mu}(\bm{s}_{1:t})$ and $\bm{\sigma}_t = Enc_{\sigma}(\bm{s}_{1:t})$. $Enc_{\mu}(\cdot)$ and $Enc_{\sigma}(\cdot)$ are Transformer encoders used to compute the mean and standard deviation of the Gaussian distribution for sequence representation, respectively.

The KL divergence between the estimated posterior and the Gaussian mixture prior cannot be directly computed analytically like the standard Gaussian prior. Therefore, similar to \cite{GMVAE-blog}, we approximate the KL divergence using the following equation:
\begin{align}
\nonumber
\mathcal{T}_{KL}^{SGM} & \approx \sum\nolimits_{t=1}^{T}(\log q_{\phi^\prime}(\bm{z}_t\mid\bm{s}_{1:t}) - \log p(\bm{z}_t)) \\
& =  \sum\nolimits_{t=1}^{T}(\log \mathcal{N}(\bm{z}_{t}\mid\bm{\mu}_t, \bm{\sigma}^{2}_t\bm{I}) - \log\sum\nolimits_{i=1}^k \alpha_t^i  \mathcal{N}(\bm{z}_t\mid \bm{m}_t^i, {\bm{\omega}_t^{i}}^2\bm{I})),
\end{align}
where $\alpha_t^i$ is the interest intensity learned from $\bm{s}_{1:t}$ and introduced in Section \ref{sec:mie}.

\vspace{-0.2cm}\subsubsection{Reconstruction Term} Since the task of SR is to predict the next item of interest, the generative process is as follows. We first sample an embedding from the estimated posterior probability $q_{\phi^\prime}(\bm{z}_t\mid\bm{s}_{1:t})\sim\mathcal{N}(\bm{\mu}_t, \bm{\sigma}_t^2\bm{I})$ using the reparameterization trick, where $\bm{z}_t = \bm{\mu}_t + \bm{\sigma}_t\odot\bm{\epsilon}$, $\bm{\epsilon}\sim\mathcal{N}(\bm{0}, \bm{I})$ is sampled from a standard Gaussian distribution. Next, we construct a decoder based on the Transformer that is almost identical to the encoder structure  to predict $p(v_{t+1}\mid \bm{z}_t)$. Let $\bm{\hat{u}}_t$ denote the output of the decoder at position $t$. Regarding the next-item prediction problem as a multi-classification problem, we calculate the interaction score between $\bm{\hat{u}}_t$ and each candidate item $v\in\mathcal{V}$ sequentially and convert it into an interaction probability using the softmax function. The $p(v_{t+1}\mid \bm{z}_t)$ is calculated as follows:

\begin{equation}
p(v_{t+1}=v\mid \bm{z}_t) = \frac{\exp(\bm{\hat{u}}_{t}^{ T}\bm{v}/\tau)}{\sum\nolimits_{v^\prime\in\mathcal{V}}\exp(\bm{\hat{u}}_{t}^{ T}\bm{v}^\prime/\tau)}.
\end{equation}

Then, we optimize the next-item prediction task by a reconstruction term as follows:
\begin{equation} 
\mathcal{T}_{recon}^{SGM} =  \sum\nolimits_{t=0}^{T-1}\mathbb{E}_{q_{\phi^\prime}(\bm{z}_t \mid \bm{s}_{1:t})} \log [p(v_{t+1} \mid \bm{z}_t)].
\end{equation}

\vspace{-0.2cm} \subsection{Training and Inference}

During training, both the SGM-VAE and MIE-VAE are simultaneously fed with interaction sequences. We jointly the two VAEs by constructing the following loss function:

\begin{align}
    \nonumber
    \mathcal{L} = -\underbrace{(\mathcal{T}_{recon}^{SGM} - \lambda\mathcal{T}_{KL}^{SGM})}_{\text{ELBO of SGM-VAE}}  - \beta_1\underbrace{(\mathcal{T}_{recon}^{MIE} - \mathcal{T}_{KL}^{MIE})}_{\text {ELBO of MIE-VAE}} + \underbrace{\beta_2\mathcal{L}_{orth}}_{\text {Regularization}},
\end{align}
where $\lambda, \beta_1, \beta_2$ are hyperparameters. $\lambda$ is introduced as weight for the KL term to avoid posterior collapse and over-regularization \cite{RecVAE}.

As for inference, we learn the posterior distribution $\mathcal{N}(\bm{\mu}_t, \bm{\sigma}_t^2\bm{I})$ of input sequences with SGM-VAE first. Then, we use $\bm{\mu}_t$ to calculate the user representation $\bm{\hat{u}}_t$ and generate the final recommendation.

\section{Experiments}





\begin{table}[t]
    \centering
    \caption{Statistics of datasets.}
    \scriptsize
    \setlength{\tabcolsep}{2mm}
    \label{tab:datasets}
    \begin{tabular}{c|ccccc}
    \toprule Dataset & \#Users & \#Items & \#Interactions & Avg. seq. len. \\
    \midrule Beauty & 22,363 & 12,101 & 198,502 & 8.3 \\
    Toys & 19,412 & 11,924 & 167,597  & 8.6 \\
    Office & 4,905 & 2,420 & 53,258 & 10.8 \\
    MovieLens & 6,040 & 3,707 & 1,000,209 & 165.6 \\
    \bottomrule
    \end{tabular}
\end{table}

\vspace{-0.2cm} \subsection{Experimental Setting} 

\vspace{-0.2cm}\subsubsection{Datasets} We adopt four publicly available datasets, including three Amazon datasets, namely \textit{Beauty}, \textit{Toys and Games}  (\textit{Toys} for short), and \textit{Office Product}  (\textit{Office} for short), and \textit{MovieLens}.  We list the statistics of these datasets in Table \ref{tab:datasets}. Following common settings in previous settings \cite{Re4}, we treat all rating behaviors in Amazon datasets as implicit feedback and ratings with 4 and 5 scores in MovieLens as positive feedback. Moreover, we eliminate users and items that have less than five interaction records. Then, we organize interactions in each dataset into user interaction sequences in ascending order based on their timestamps.  For each user, the most recently interacted item is used for testing, the second most recent item is used for validation, and the rest of the interacted items are used for training. We leverage the three Amazon datasets for performance comparison to other recommendation models and employ MovieLens to analyse recommendation diversity.

\vspace{-0.2cm}\subsubsection{Metrics} Recall@K and NDCG@K are used as metrics to evaluate the performance of the recommendation results. We report the results where $K=20$ and  $K=40$.

\vspace{-0.2cm}\subsubsection{Compared Models} 

We compare SIGMA to several representative models, which can be separated into four classes. 1) Collaborative filtering: \textit{BPRMF} \cite{MFBPR} and \textit{LightGCN} \cite{LightGCN}. 2) Attention-based SR models: \textit{SASRec} \cite{SASRec} and \textit{Bert4Rec} \cite{BERT4Rec}. 3) VAE-based SR models: \textit{SVAE} \cite{SVAE}, \textit{ACVAE} \cite{ACVAE} and \textit{DT4SR} \cite{DT4SR}. 4)  Multi-interest recommendation models: \textit{MIND} \cite{MIND}, \textit{ComiDR} and \textit{ComiSA}\cite{ComiRec}.

\vspace{-0.2cm}\subsubsection{Implementation Details} We implement SIGMA using PyTorch and conduct experiments on NVIDIA GeForce RTX 3090. The number of categories is tuned in [2, 4, 8, 16], and ultimately set to 4 for the three Amazon datasets and 8 for the MovieLens dataset. The $\lambda$ searched among [0.01, 0.001, 0.0001, 0.00001], and finally set to 0.0001. Both $\beta_1$ and $\beta_2$ are also searched among [0.01, 0.001, 0.0001, 0.00001] and set to 0.01.  To prevent over-fitting, we adopt an early stop strategy to terminate experiments when there is no improvement in Recall@20 after 100 epochs. The maximum sequence length is set to 100. Sequences longer than 100 are truncated, while sequences shorter than 100 are padded with item elements at the front until they reach a length of 100. The learning rate is set as 0.001, the hidden dimension is set as 128, the model dropout probability is set as 0.3, and the number of attention heads is set as 4. For fairness, the embedding dimension, batch size, maximum sequence length, and hyperparameters in attention mechanisms are set consistently among all competitors. The source code is available at https://anonymous.4open.science/r/SIGMA-1E6D.


%

\vspace{-0.2cm} \subsection{Performance Comparison}

\begin{table*}[t]
    \caption{Recommendation performance on three datasets, where R denotes Recall and N denotes NDCG. We bold the best results and underline the second best results of each
metric. The last column is the relative improvements compared with the best baseline results.}
    \label{tab:performance-comparison}
    \centering
    \tiny
    \begin{tabular}{c|cccc|cccc|cccc}
    \toprule 
    & \multicolumn{4}{c|}{\textbf{Beauty}} &  \multicolumn{4}{c|}{\textbf{Office}}  &  \multicolumn{4}{c}{\textbf{Toys}} \\
    \midrule
    Models & R@20 & R@40 & N@20 & N@40 & R@20 & R@40 & N@20 & N@40 & R@20 & R@40 & N@20 & N@40 \\
    \midrule
    BPRMF & 0.0739 & 0.1089 & 0.0311 & 0.0383 & 0.0483 & 0.0718 & 0.0218 & 0.0266 & 0.0692 & 0.1007 & 0.0304 & 0.0369 \\
    LightGCN & 0.0759 & 0.1112 & 0.0306 & 0.0378 & 0.0532 & 0.0797 & 0.0243 & 0.0297 & 0.0671 & 0.0977 & 0.0287 & 0.0349 \\
    \midrule
    Bert4Rec & 0.0890 & 0.1285 & 0.0395 & 0.0476 & 0.1350 & 0.2230 & 0.0551 & 0.0729 & 0.0699 & 0.0982 & 0.0318 & 0.0376 \\
    SASRec & 0.0952 & 0.1389 & 0.0420 & 0.0509 & \underline{0.1478} & \underline{0.2251} & \underline{0.0657} & \underline{0.0815} & 0.1112 & \underline{0.1479} & \underline{0.0539} & \underline{0.0614} \\
    \midrule
    SVAE & 0.0268 & 0.0417 & 0.0102 & 0.0132 & 0.0988 & 0.1647 & 0.0389 & 0.0523 & 0.0178 & 0.0260 & 0.0069 & 0.0086 \\
    ACVAE & 0.0951 & 0.1294 & \underline{0.0467} & \underline{0.0537} & 0.1327 & 0.2075 & 0.0560 & 0.0713 & 0.0722 & 0.1030 & 0.0359 & 0.0421 \\
    DT4SR & \underline{0.0982} & \underline{0.1404} & 0.0446 & 0.0533 & 0.1429 & 0.2186 & 0.0643 & 0.0797 & \underline{0.1130} & 0.1478 & 0.0515 & 0.0560 \\
    \midrule
    MIND & 0.0588 & 0.0859 & 0.0247 & 0.0297 & 0.0614 & 0.1066 & 0.0240 & 0.0354 & 0.0555 & 0.0801 & 0.0243 & 0.0289 \\
    ComiRec-DR & 0.0373 & 0.0583 & 0.0148 & 0.0194 & 0.0263 & 0.0601 & 0.0106 & 0.0178 & 0.0180 & 0.0266 & 0.0080 & 0.0097 \\
    ComiRec-SA & 0.0574 & 0.0906 & 0.0200 & 0.0271 & 0.0685 & 0.1193 & 0.0233 & 0.0335 & 0.0474 & 0.0722 & 0.0179 & 0.0228 \\
    \midrule
    SIGMA & \textbf{0.1048} & \textbf{0.1494} & \textbf{0.0477} & \textbf{0.0568} & \textbf{0.1615} & \textbf{0.2334} & \textbf{0.0722} & \textbf{0.0869} & \textbf{0.1151} & \textbf{0.1510} & \textbf{0.0549} & \textbf{0.0629} \\
    Improv. & 6.69\% & 6.41\% & 2.13\% & 5.77\% & 9.25\% & 3.70\% & 9.95\% & 6.61\% & 1.86\% & 2.09\% & 1.94\% & 2.38\% \\
    \bottomrule
    \end{tabular}
\end{table*}

The experimental results are shown in Table \ref{tab:performance-comparison}, from which we have the following observations.

\textit{Firstly, the quality of posterior probability estimation plays a crucial role in determining the recommendation performance of VAE-based SR models}. When compared to other competitors, such as ACVAE and DT4SR, SVAE exhibits extremely poor performance. This can be attributed to the fact that SVAE adheres to the conventional VAE approach, which assumes that the embeddings of each user are sampled from a standard Gaussian distribution. The similarity between the assumed prior distribution and the estimated posterior distribution results in the collapse of user representations, leading to the loss of personalized information. This phenomenon is particularly pronounced when the number of items is large, as observed in datasets such as Beauty and Toy. ACVAE and DT4SR, which enhance the inference of latent variables by employing the adversarial variational Bayes technique and utilizing Elliptical Gaussian distributions, respectively, achieve significant performance improvement compared to SVAE.

 \textit{Moreover, existing multi-interest recommendation  models that acquire multiple interest representations for users have not exhibited superior performance}. This discrepancy in the expectation arises from two main factors. Firstly, unlike other models that solely prioritize recommendation accuracy, multi-interest recommendation models also need to take into account the diversity of recommendations. Consequently, these two objectives may lead to an optimization dilemma. Secondly, multi-interest models constructed with dynamic routing, such as MIND and ComiRec-DR, do not consider the sequential nature of a user's history interactions. Instead, they treat these interactions as an unordered item set.

\textit{Finally, the proposed model SIGMA  outperforms all competitors,  highlighting its effectiveness}. This superiority can be attributed to the semantic and expressive sequence representation learned by SIGMA. Notably, SIGMA demonstrates the greatest improvement on the Office dataset that comprises longer average interaction sequences than those of the other two datasets, which means that SIGMA is more adept at learning high-quality interest representations from longer sequences

\begin{table}[t]
    \centering
    \setlength{\tabcolsep}{2mm}
    \caption{Effect of MIE-VAE, where $\lambda^\prime$ denotes the weight of KL-divergence term in the model variant $\text{SIGMA}_{uni}$. }
    \label{tab:ablation-mievae}
    \scriptsize
    \begin{tabular}{c|c|cccc}
        \toprule
        & Datasets & Recall@40 & NDCG@40 \\ 
        \midrule
        \multirow{3}{*}{$\lambda^\prime = 1$} & Beauty &  0.0620 $\downarrow$ \textbf{58.52\%} & 0.0203 $\downarrow$ \textbf{64.31\%} \\
        & Office & 0.1784 $\downarrow$ \textbf{23.58\%} & 0.0554  $\downarrow$ \textbf{36.20\%} \\
        & Toy & 0.0604 $\downarrow$ \textbf{60.01\%} & 0.0194 $\downarrow$ \textbf{69.21\%} \\
        \midrule
        \multirow{3}{*}{$\lambda^\prime = 0.0001$} & Beauty & 0.1333 $\downarrow$ \textbf{10.75\%} & 0.0523 $\downarrow$ \textbf{7.99\%} \\
        & Office & 0.2243 $\downarrow$ \textbf{3.93\%} &  0.0828 $\downarrow$ \textbf{4.75\%}\\
        & Toy & 0.1439 $\downarrow$ \textbf{4.67\%} & 0.0601 $\downarrow$ \textbf{4.34\%}\\
        \bottomrule
    \end{tabular}
    \label{tab:array}
\end{table}

\vspace{-0.2cm} \subsection{Ablation Study}

\vspace{-0.1cm}\subsubsection{Effect of MIE-VAE} In this section, we investigate the role of the MIE-VAE. A model variant of SIGMA is constructed by removing the MIE-VAE, i.e., setting the prior as a standard Gaussian distribution rather than the mixture of interest distributions learned with MIE-VAE, which is denoted as $\text{SIGMA}_{uni}$. The weight  of the KL-divergence term in $\text{SIGMA}_{uni}$, referred to as $\lambda^\prime$, is set in two ways: 1) Following SVAE, $\lambda^\prime$ is set to 1. 2) Similar to $\lambda$ in the proposed SIGMA, the optimal value of $\lambda^\prime$ is searched for among $[0.01, 0.001, 0.0001, 0.00001]$, and finally it is set to 0.0001. 

The experimental results, displayed in Table \ref{tab:ablation-mievae}, demonstrate a significant decline in recommendation performance when $\lambda^\prime = 1$, with some metrics even exhibiting a drop of over 50\% compared to SIGMA. This shows that employing plain priors to optimize estimated posterior will lead to posterior collapse, which adversely affects the final performance. However, it still performs slightly better than SVAE in Table \ref{tab:performance-comparison}, mainly due to the strong sequence representation capability of Transformer compared to RNN. On the other hand, when $\lambda^\prime = 0.0001$, the experimental results show a substantial improvement compared to the case when $\lambda^\prime = 1$, but there is still a significant gap compared to SIGMA. For example, on Beauty, Recall@40, and NDCG@40 experience a decline of 10.75\% and 7.99\%, respectively. This demonstrates the advantageous contribution of the Gaussian Mixture prior based on multiple interests to the overall recommendation performance.

\vspace{-0.2cm}\subsubsection{Effect of SGM-VAE }
In SIGMA, the sequence representations learned by SGM-VAE adeptly integrate users' diverse interests into user representation and also account for the varying intensities of these interests. To assess the effect of SGM-VAE, we train MIE-VAE separately and use it to generate recommendation results. We retrieve 40 items for each user interest, and then select the top 40 items with the highest matching scores from the pool of all matched items as the final recommendations. The experimental results, as depicted in Table \ref{tab:abl_sgm}, show a significant performance decline, which clearly illustrates the substantial contribution of SGM-VAE. This drop in performance is likely due to the fact that MIE-VAE treats all user interests with equal importance, whereas SGM-VAE differentiates between the intensities of different user interests within its prior distribution.

\begin{table}[t]
    \centering
    \setlength{\tabcolsep}{2mm}
    \caption{Effect of SGM-VAE. }
    \label{tab:abl_sgm}        
    \scriptsize
    \begin{tabular}{c|c|c|c}
        \toprule
        Metrics & Beauty & Office & Toy \\ 
        \midrule
        Recall@20 &  0.0890 $\downarrow$ \textbf{15.09\%}  & 0.1150 $\downarrow$ \textbf{28.80\%}  & 0.0978 $\downarrow$ \textbf{15.01\%}\\
        Recall@40 & 0.1315	$\downarrow$ \textbf{12.00\%} & 0.1806 $\downarrow$ \textbf{22.61\%} & 	0.1319 $\downarrow$ \textbf{12.66\%}	 \\
        NDCG@20 &  0.0383	$\downarrow$ \textbf{19.78\%} & 0.0517 $\downarrow$ \textbf{28.35\%}  & 0.0482 $\downarrow$ \textbf{12.16\%}\\
        NDCG@40 &  0.0469 $\downarrow$ \textbf{17.40\%} & 0.0651 $\downarrow$ \textbf{25.09\%} & 0.0552 $\downarrow$ \textbf{12.28\%}\\
        \bottomrule
    \end{tabular}
\end{table}

\vspace{-0.2cm}\subsubsection{Effect of Orthogonal Constrain}

\begin{table}[t]
    \centering
    \scriptsize
    \setlength{\tabcolsep}{2mm}
    \caption{Effect  of the orthogonal constrain, i.e., $\mathcal{L}_{orth}$ }
    \label{tab:study_orth}
    \begin{tabular}{c|c|c|c}
        \toprule
        Metrics & Beauty & Office & Toy \\ 
        \midrule
        Recall@20 &  0.0987	$\downarrow$ \textbf{5.83\%} & 0.1201	$\downarrow$ \textbf{25.65\%} & 0.1046	$\downarrow$ \textbf{9.10\%} \\
        Recall@40 & 0.1415	$\downarrow$ \textbf{5.27\%} & 0.1827	$\downarrow$ \textbf{21.73\%} & 0.1455	$\downarrow$ \textbf{3.62\%}	 \\
        NDCG@20 &  0.0430	$\downarrow$ \textbf{9.85\%} & 0.0512	$\downarrow$ \textbf{29.07\%} & 0.0495	$\downarrow$ \textbf{9.81\%} \\
        NDCG@40 &  0.0517	$\downarrow$ \textbf{8.97\%} & 0.0640    $\downarrow$ \textbf{26.38\%} & 0.0579   $\downarrow$ \textbf{8.02\%} \\
        \bottomrule
    \end{tabular}
\end{table}

We construct a model variant by removing the orthogonal constrain of category embeddings, i.e., $\mathcal{L}_{orth}$, to investigate the effect of the orthogonal loss. Its performance is shown in Table \ref{tab:study_orth}. It can be observed that without $\mathcal{L}_{orth}$, there is a significant decrease in recommendation accuracy, especially on the Office dataset. The experimental results confirm the necessity of the orthogonal constrain. This is because the orthogonal constrain of interest embeddings helps in learning semantic category representations, which in turn benefits the multi-interest priors and enhances the quality of sequence representations and recommendation.

\vspace{-0.2cm} \subsection{Study about Gaussian Mixture Prior} 

\vspace{-0.2cm}\subsubsection{Impact on Recommendation Diversity}

In this section, we explore the effectiveness of the multimodal Gaussian prior in capturing users' diverse interests and improving the diversity of recommendations. For a given user $ u $, we assume the $K$ recommended items for user $u$ to be $ [\hat{v}_{u1}, \hat{v}_{u2}, \ldots, \hat{v}_{uK}] $. Similar to ComiRec \cite{ComiRec},  we introduce the following metric to measure the diversity of these recommendations:
\begin{equation}
    Diversity@K=\frac{1}{|\mathcal{U}|}\sum\nolimits_{u\in\mathcal{U}}\frac{\sum\nolimits_{j=1}^K\sum\nolimits_{k=j+1}^K\delta(C(\hat{v}_{uj})\neq C(\hat{v}_{uk}))}{K\times(K-1)/2},
\end{equation}
where $C(\hat{v}_{ui})$ represents the category of the $ i $-th recommended item for user $ u $, $\delta(\cdot)$ is an indicate function. We conduct experiments on the MovieLens dataset, where each movie can be associated with multiple genres. We categorize two movies as being in the same category if they share at least one genre in common. 

\begin{table}[t]
    \centering
    \setlength{\tabcolsep}{2mm}
    \caption{Impact of the Gaussian mixture prior on recommendation diversity. }
    \label{tab:ablation}    
    \scriptsize
    \begin{tabular}{c|ccc|ccc}
        \toprule
    & \multicolumn{3}{c|}{@10} &  \multicolumn{3}{c}{@40} \\
        \midrule
        $\lambda$ & Recall & NDCG & Diversity & Recall & NDCG & Diversity \\ 
        \midrule
0.01	& 0.1245	& 0.0617	& \textbf{0.2645}	& 0.3195	& 0.1055	& \textbf{0.3934} \\
0.001	& 0.1311	& 0.0638	& 0.2443	& 0.3145	& 0.1049	& 0.3878 \\
0.0001	& \textbf{0.1326}	& \textbf{0.0651}	& 0.2424	& \textbf{0.3228}	& \textbf{0.1074}	& 0.3841 \\
        \bottomrule
    \end{tabular}
    \label{tab:diversity}
\end{table}

We vary the weight of the KL-divergence term between the posterior of sequence representation and the semantic Gaussian mixture prior, i.e., $ \lambda $, at levels of 0.01, 0.001, and 0.0001, and document both the accuracy and diversity of the recommendation. The results, presented in Table \ref{tab:diversity}, reveal that as $\lambda$ increases, the diversity of the recommendations also increases, albeit with a slight decrease in accuracy. This suggests that incorporating a multimodal prior distribution into the sequential representation can effectively enhance the diversity of recommendations. By choosing an appropriate  weight $ \lambda $, it is possible to strike a balance between accuracy and diversity in the recommendation outcomes.

\vspace{-0.2cm}\subsubsection{Impact of Category Quantity} 
\begin{figure*}[!t]
    \centering
    \subfigure[On the Beauty dataset.]{ 
        \begin{minipage}[b]{0.3\textwidth} 
            \centering 
            \includegraphics[width=\textwidth]{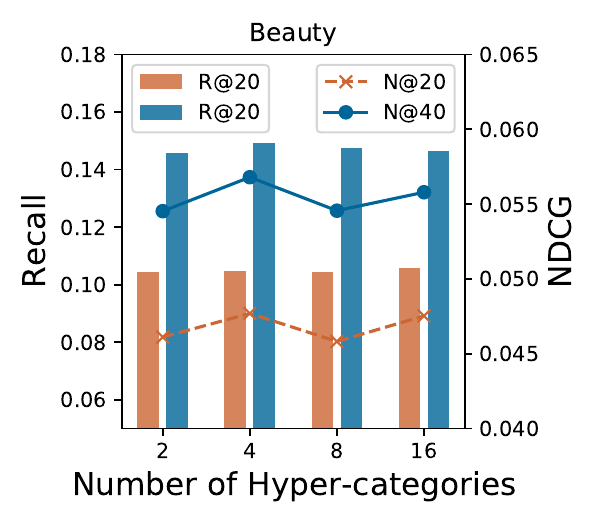} 
    \end{minipage}}%
    \subfigure[On the Office dataset.]{ 
        \begin{minipage}[b]{0.3\textwidth} 
            \centering 
            \includegraphics[width=\textwidth]{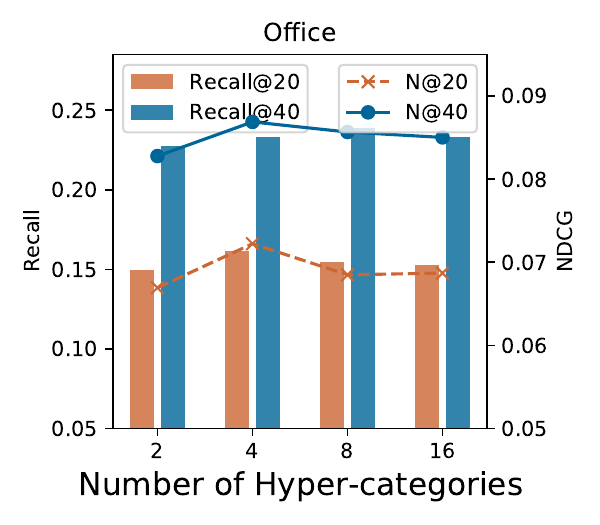} 
    \end{minipage}}%
    \subfigure[On the Toy dataset.]{ 
        \begin{minipage}[b]{0.3\textwidth} 
            \centering 
            \includegraphics[width=\textwidth]{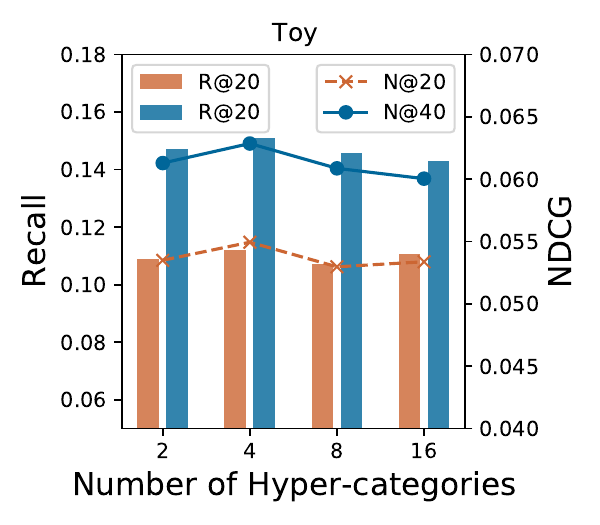} 
    \end{minipage}}%
    \caption{ Impact of category quantity.} 
    \label{fig:impact-hypercategories} 
\end{figure*}

The number of categories $k$ affects the multiple interest extraction for each user, which further affects the recommendation performance of SIGMA. We evaluate the performance of SIGMA with different category quantities: 2, 4, 8, and 16. As shown in Figure \ref{fig:impact-hypercategories}, the model performs best when $k=4$ on both of the two datasets. This is because a small number of categories cannot effectively partition candidate items, enhancing the possibility of modeling completely unrelated items within the same user interest. On the other hand, a large number of categories may separate correlated items into different interests, impacting the accuracy of interest extraction, especially for short sequences.

\section{Related Work}

\vspace{-0.2cm} \subsection{VAE for Recommendation}

 Mult-VAE \cite{MultiVAE} is the pioneering work to introduce VAE into collaborative filtering, which assumes that the latent representation of each user is sampled from a standard Gaussian distribution. RecVAE \cite{RecVAE} proposes a new composite prior for training by alternating updates to enhance the performance of Mult-VAE. SVAE \cite{SVAE} and VSAN \cite{VSAN} adopt VAE for SR using recurrent neural network and self-attention as encoder and decoder, respectively. To further enhance the quality of the inferred latent variables, ACVAE \cite{ACVAE} introduces adversarial learning via AVB framework to the sequential recommendation, which reduces the relevance between different dimensions in latent variables. DT4SR \cite{DT4SR} adopts Elliptical Gaussian distributions to describe items and sequences with uncertainty. ContrastVAE \cite{ContrastVAE} extends conventional single-view ELBO to two-view case and naturally incorporates contrastive learning to the framework of VAE-based SR. However, existing VAE-based SR models assume a unimodal Gaussian latent space for sequence representation, which is difficult to describe complex distributions and limits the representation power of the hidden space. 

\vspace{-0.2cm} \subsection{Multi-interest Recommendation} 

 Multi-interest recommendation models learn multiple different interest embeddings from history interaction sequences. For example, SASRec \cite{SASRec}, SDM \cite{lv_sdm_2019} and ComiRec-SA \cite{cen_controllable_2020} propose to apply the multi-head attention mechanism to learn the different interests of users. MIND \cite{MIND} and ComiRec-DR \cite{cen_controllable_2020} design dynamic routing \cite{sabour2017dynamic}, i.e., variants of the capsule network, treating high-level capsules as interests obtained by item clustering. Octopus \cite{liu_octopus_2020} builds an elastic archive network to recognize multiple interests of users. After exacting multiple interest vectors, these multi-interest recommendation models produce one-way recommendation results with each extracted interest embedding and aggregate them to form final recommendation by choosing the overall top-k items with the maximal matching scores. However, due to that they learn only determinate interest representations, they are easily affected by data noise and sparsity. 

\section{Conclusion}
In this paper, we propose adopting a Gaussian mixture prior distribution to describe user interests implied in each interaction sequence and build a novel variational autoencoder for SR, named SIGMA. Specifically, we construct a multi-interest extraction VAE that learns a unimodal Gaussian distribution for each interest and mixes them based on the interest intensity. By incorporating a KL divergence between the estimated posterior distribution of sequence representation and the  Gaussian mixture distribution composed of multiple interests, SIGMA enables the sequence representation to capture the complex and diverse preferences of users. Experimental results show that SIGMA surpasses existing SR methods, demonstrating the contribution of the semantic Gaussian mixture prior to performance improvement. 

%
%
%
\bibliographystyle{splncs04}
\bibliography{ref}
\end{document}